\documentclass[twocolumn,preprintnumbers,superscriptaddress,amsmath,amssymb,letter]{revtex4}

\usepackage{lineno}
\usepackage{dcolumn}

\usepackage{bm}
\usepackage{amsmath}
\usepackage{multirow}
\usepackage{graphicx}
\usepackage{float}
\usepackage[colorlinks=true]{hyperref}

\begin{document}

\title{Quantum mechanical formulation of the Busch theorem}

\author{K. Floettmann}

\email{Klaus.Floettmann@DESY.de}

\affiliation{Deutsches Elektronen-Synchrotron, Notkestra\ss e 85, 22607 Hamburg, Germany}

\author{D. Karlovets}

\email{d.karlovets@gmail.com}

\affiliation{Tomsk State University, Lenina Ave.\,36, 634050 Tomsk, Russia}

\date{\today}

\begin{abstract}
Due to the conservation of the canonical angular momentum charged particle beams which are generated inside a solenoid field acquire a kinetic angular momentum outside of the solenoid field. The relation of kinetic orbital angular momentum to the field strength and the beam size on the cathode is called Busch theorem. We formulate the Busch theorem in quantum mechanical form and discuss the generation of quantized vortex beams, i.e., beams carrying a quantized orbital angular momentum.
Immersing a cathode in a solenoid field presents an efficient and flexible method for the generation of electron vortex beams, while, e.g., vortex ions can be generated by immersing a charge stripping foil in a solenoid field. Both techniques are utilized at accelerators for the production of non-quantized vortex beams. As highly relevant use case we discuss in detail the conditions for the generation of quantized vortex beams from an immersed cathode in an electron microscope. General possibilities of this technique for the production of vortex beams of other charged particles are pointed out.
\end{abstract}


\maketitle
{\it Introduction. --}
The generation of vortex beams as twisted photons~\cite{Allen1992}, vortex neutrons~\cite{Clark2015}, or vortex electrons has inspired versatile theoretical studies and interesting experiments or proposals to unveil the basic properties of such beams and of effects of quantum interference and coherence in particle collisions, inaccessible with ordinary beams \cite{Ivanov12, DK2017, Schattschneider2017, DK2020, Ivanov2020_1, Ivanov2020_2, PRC2019}. Quantized vortex electrons -- i.e., electron beams carrying a quantized orbital angular momentum (OAM) -- generated in electron microscopes~\cite{Uchida2010, Verbeeck2010, McMorran2011} can be applied as probes for the study of chiral~\cite{Juchtmans2015} or magnetic structures~\cite{Lloyd2012}, and enables magnetic mapping with atomic resolution~\cite{Schattschneider2012}.

The electron microscope community devised several methods to produce and analyze electron vortex beams (for a review see~\cite{Schattschneider2017}), e.g., by means of spiral phase plates~\cite{Uchida2010}, holographic diffraction gratings~\cite{Verbeeck2010}, or by the interaction with a magnetic needle, which mimics an approximate magnetic monopole~\cite{Beche2014}. The low efficiency and the limited flexibility of these methods hampers however the broad application of vortex beams for explorations of the atomic structure of matter and of fundamental interactions beyond a plane-wave approximation \cite{Ivanov12, DK2017, Schattschneider2017, DK2020, Ivanov2020_1, Ivanov2020_2, PRC2019}.

Modern accelerators also make use of beams carrying a non-quantized orbital angular momentum, predominantly for the manipulation of the phase-space of a beam~\cite{Burov2000, Burov2002, Kim2003, Sun2004}, with the aim to redistribute the phase-space volume between transverse degrees of freedom. The prevailing technique for the generation of vortex electron beams applied in accelerators makes use of a cathode which is immersed in a solenoid field. This immersion of the cathode changes the dynamics of charged particle beams in a very fundamental way. While the angular momentum, which a solenoid imparts onto a beam, is canceled exactly when the beam travels through the complete solenoid, this is not the case when the beam is generated inside the field. In simple words, the beam sees only half of a solenoid in the second case and thus a freely propagating beam with intrinsic angular momentum is generated. The relation between the solenoid field strength and the beam size on the cathode and the angular momentum of the freely propagating beam outside of the solenoid is described by the so-called Busch theorem~\cite{Busch1926}. The beams created with the use of an immersed cathode carry typically a large average angular momentum with a broad spectral distribution, or OAM bandwidth. 

Importantly, this technique can be adapted to the generation of vortex beams of all kind of charged particles. Besides electron vortex beams, ion beams carrying a non-quantized OAM have for example already been produced~\cite{Groening2017}.

In this Letter, we formulate the Busch theorem in quantum mechanical form  and discuss as use case the immersed cathode technique for the generation of vortex beams with a quantized OAM in an electron microscope. While applications in a microscope do not necessarily require the generation of pure modes with a well-defined angular momentum and a vanishing OAM quantum uncertainty (OAM bandwidth), the discussion concentrates on this operation mode to point out the most stringent requirements.\\General possibilities of this technique for the production of vortex beams of other charged particles are pointed out.\\
 
{\it Classical particle in a solenoid. --}
Solenoid fields can -- as any rotational symmetric electromagnetic field -- be developed in the form of a polynomial series as:
	\begin{equation}
	\begin{gathered}
  {B_z}(z,r) = {B_{z,0}} - \frac{{{r^2}}}{4}B_z^{''} + \frac{{{r^4}}}{{64}}B_z^{''''}... \hfill \\
  {B_r}(z,r) =  - \frac{r}{2}B_z^{'} + \frac{r^3}{16}B_z^{'''} - \frac{{{r^5}}}{{384}}B_z^{'''''}... \hfill \\
  {B_\theta } = 0. \hfill 
\end{gathered}
\label{Eq1.1}
\end{equation}
Here, ${B_z}$, ${B_r}$ and ${B_\theta }$ are the field components in a cylindrical system with the coordinates $\{r, \theta, z\}$, ${B_{z,0}}$ denotes the on-axis field and a prime indicates a derivative with respect to the longitudinal axis $z$.\\
The transverse fields of solenoids installed in electron microscopes and accelerators have to be linear within the radius of the total beam size, so as to preserve the beam quality. The discussion will thus concentrate in the following on the first terms of Eqs~\ref{Eq1.1}.\\
The transverse canonical momenta in Cartesian coordinates $\{x, y, z\}$ are defined as:
	\begin{equation}
	\begin{gathered}
  {{\tilde p}_x} = {p_x} + q {A_x} \cong {p_x} - \frac{{q {B_{z,0}}}}{2}y \hfill \\ 
  {{\tilde p}_y} = {p_y} + q {A_y} \cong {p_y} + \frac{{q {B_{z,0}}}}{2}x, \hfill
\end{gathered}
\label{Eq1.2}
\end{equation}
where the charge $q$ can be of arbitrary sign, ${p_x}$ and ${p_y}$ denote the mechanical momenta of a particle, and ${A_x}$ and ${A_y}$ stand for the vector potential of the field, which is in linear approximation ${A_\theta } \cong \frac{r}{2}{B_{z,0}}$ (cf.~\ref{Eq1.1}). 	

The canonical angular momentum follows as:
\begin{equation}
\begin{gathered}
  \tilde L = x{{\tilde p}_y} - y{{\tilde p}_x} = r{p_{_\theta }} + q r{A_\theta } \\ 
   \approx x{p_y} - y{p_x} + \frac{{q {B_z}}}{2}\left( {{x^2} + {y^2}} \right) \\ 
\end{gathered}
\label{Eq1.3} 
\end{equation}
The canonical angular momentum is a conserved quantity of motion~\cite{Brillouin1945}. Note that Eq.~\ref{Eq1.3} is defined with respect to the axis of the solenoid field, which is not the axis of the spiral motion of a particle in an electron microscope or accelerator configuration. Particles entering a solenoid with negligible transverse momentum components and transverse offset $r$ relative to the solenoid axis, will build up an azimuthal momentum in the fringe field of the solenoid such that they rotate with the radius $r/2$ around an axis, which has an offset of $r/2$ to the solenoid axis. Thus all particles will cross the axis of the solenoid after half a period and the beam is being focused.

The vector potential is related to the magnetic flux in a circle of the radius $r$ by:
\begin{equation}
	{A_\theta } = \frac{1}{{2\pi }}\oint {{A_\theta }dl}  = \frac{1}{{2\pi }}\int {\left( {\triangledown  \times A} \right)} \;ds = \frac{1}{{2\pi }}\int {Bds = \frac{\Phi }{{2\pi }}}, 
\label{Eq1.4}
\end{equation}
where the integrals cover the circumference and the area of the circle described by $r$. Thus
	\begin{equation}
	\tilde L = r{p_\theta } + \frac{{q \Phi }}{{2\pi }}
	\label{Eq1.5}
	\end{equation}
In this form, the conservation of the canonical angular momentum is called Busch theorem \cite{Busch1926}.

The Busch theorem implies the conservation of the mechanical angular momentum $L$ for the complete passage through a solenoid, 
because outside of a solenoid $\tilde L = L=xp_y-yp_x$ holds, i.e., if $L = 0$ when the beam enters the solenoid, it will exit also with $L= 0$. 
However, when for example a cathode is immersed in a solenoid field electrons starting with $ L= 0$ on the cathode carry the canonical angular momentum of $\tilde L =  \frac{{q\Phi }}{{2\pi }}$, 
which turns into the kinetic angular momentum $r{p_\theta } =  \frac{{q\Phi }}{{2\pi }}$ in the field free region.

This transfer is entirely due to the Lorentz force of the radial field components. In the idealized case of a long constant solenoid field and no additional forces, the particles would move on straight lines parallel to the solenoid field lines until the fringe field region is reached. Neither in the case of the immersed cathode nor in the case of the non-immersed cathode, does the beam trajectory encircle the area which enters the integrals in Eq.~\ref{Eq1.4}.\\

{\it Quantum particle in a solenoid. --} Let us now consider the quantum wave packet of a charged particle which is created inside a solenoid field with a vanishing kinetic angular momentum. Following Dirac~\cite{Dirac1931}, its canonical angular momentum is quantized with an integer quantum number $\ell$ as 
\begin{eqnarray}
\tilde{L} = \ell \hbar = \frac{q}{2} B_{z,0} \langle r^2 \rangle=\frac{q \langle\Phi\rangle}{2\pi}
\label{Eq1.6}
\end{eqnarray}
and so is the mean flux of the magnetic field, which reads as
\begin{eqnarray}
\langle\Phi\rangle = \left\langle \oint {\bm A} d{\bm l} \right\rangle = \frac{4\pi}{|q|}\, \hbar\, \frac{\langle r^2\rangle}{r_H^2},
\label{Eq1.6.1}
\end{eqnarray}
where $\langle ... \rangle$ denotes a quantum-mechanical averaging and $r_H^2 = 4\hbar/|q|B_{z,0}$.

Thus, the particle packet acquires a quantized orbital angular momentum when leaving the solenoid field.
Eqs~(\ref{Eq1.6}) and~(\ref{Eq1.6.1}) represent a quantum counterpart of the classical Busch theorem.
Due to similarities of the fringe fields of the solenoid and those of a tip of a magnetic needle, the effect is somewhat analogous to that arising 
from the interaction with an effective magnetic monopole~\cite{Beche2014} -- the particle's wave function gets an Aharonov-Bohm phase,
\begin{eqnarray}
\Psi \to \Psi\, \exp\left\{i\theta\frac{q}{2\pi\hbar}\left\langle \oint {\bm A} d{\bm l} \right\rangle\right\} = \Psi\, e^{i\ell\theta}.
\label{Eq1.6.2}
\end{eqnarray}
As can be seen, the sign of the resultant angular momentum $\ell$ is correlated with the sign of the particle's charge and with the sign of the magnetic flux.

For a particle with charge $|q| = |eZ|$, a convenient estimate of the intrinsic angular momentum acquired by the particle is given by:
\begin{eqnarray}
|\ell| \approx 1.5 \times 10^{-3}\, |Z| \langle r^2\rangle [\text{nm}^2]\, |B_{z,0}| [\text{T}],
\label{Eqell}
\end{eqnarray}
where $r$ is measured in nanometers and the magnetic field $B_{z,0}$ in Tesla. Formally, the estimate (\ref{Eqell}) coincides with its classical counterpart~\cite{Sun2004} 
but $r$ is the rms width of a single-particle wave packet now, not that of a classical beam of many particles.\\
\begin{figure}[ttt]
\centering
    \includegraphics*[width=80mm]{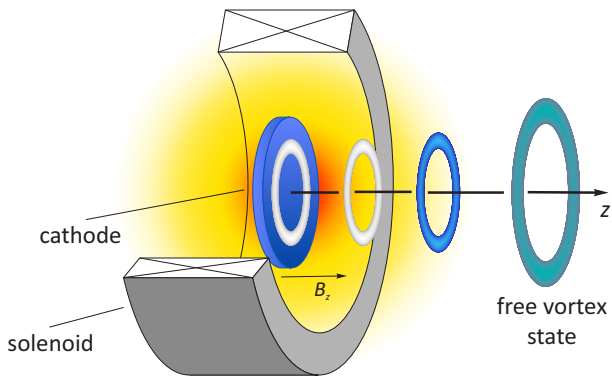}
    \caption{Illustration of the immersed cathode technique. Electrons with a ring-shaped probability density are released from a cathode which is immersed in a solenoid field. When the electrons leave the solenoid field an orbital angular momentum is imparted and a free vortex state is generated.}
\label{fig.1}
\end{figure}

{\it Case example: vortex electrons --} In accelerator physics, beams with angular momentum are generally not desirable, because the angular momentum leads to an additional beam divergence and thus the beam quality is degraded. In many particle sources it is however necessary to place the first focusing solenoid lens so close to the cathode that the fringe field extends up to and behind the cathode. These sources are equipped with a so-called bucking coil, i.e., a solenoid behind the cathode plane which is excited with opposite polarity to the focusing lens in order to compensate the longitudinal field component on the cathode. In such a configuration the angular momentum of the beam can be easily modified by adjusting the current in the bucking coil.

A combination of a focusing lens and the bucking coil was also the basis for the intensive experimental program on vortex beams at the FNAL photoinjector~\cite{Edwards2000, Edwards2001, Sun2004, Piot2006}. The experiments demonstrated the generation of angular momentum dominated beams and the repartitioning of the beam emittance in a so-called flat-beam adapter~\cite{Brinkmann2001}. A comparison of this beam adapter to the mode converter known for laser beams is discussed in a parallel publication to this report~\cite{Floettmann2020}.

In a photoinjector, the cathode consists of a photo emissive material. Electrons are released by directing a laser beam onto the cathode. Usually a uniform emission within a specified radius is aimed for, which can be described as a superposition of Laguerre-Gaussian beams~\cite{Gori1994}. Thus a corresponding broad spectrum of angular momenta is generated. In the experiments at FNAL, average angular momenta of $\sim$100 neV$\times$s have been measured~\cite{Sun2004}, corresponding to $\sim10^8 \hbar$. The rms beam size on the cathode in these experiments was in the millimeter range and a solenoid field of about 0.1~T was applied on the cathode, in accord with Eq.~(\ref{Eqell}). Beams with such a high angular momentum could hardly be handled in an electron microscope and the observation of quantum effects, e.g., discontinuous quantum steps, are rather unlikely at too high values of the OAM. Thus, a pure mode with a much lower quantum number is desirable for experimental conditions in which quantization could be observed.

According to the corresponding uncertainty relations~\cite{Carr}, the generation of a pure mode requires the emission from a ring, centered to the axis of the solenoid, as illustrated in Figure~\ref{fig.1}. This can be realized, e.g., by directing a Laguerre-Gauss mode laser onto the photo cathode immersed in the solenoid field. In case of a mode with a radial mode number $n = 0$, the mode consists of a single ring with a nearly Gaussian cross-section of the ring. Electrons would thus naturally be emitted into the transverse distribution of a Laguerre-Gauss mode. Alternatively, a micro-structured photo cathode with a ring-shaped emissive area on a non-emissive background could be envisaged. Also field emission, for example from a ring of field emitting tips, is not excluded. In the latter cases the emission would not match the transverse distribution of a Laguerre-Gauss ring; however, it is conceivable to realize even smaller structures, than by means of a focused laser beam.

Besides the spatial transverse distribution, the uncorrelated transverse beam momentum distribution also needs to be matched to the phase-space of a Laguerre-Gauss mode. The uncorrelated momentum spread of a Laguerre-Gauss mode increases with the increasing angular momentum as ${\sigma _{px}} = \hbar \frac{{\left| \ell \right| + 1}}{{2{\sigma _x}}}$, where ${\sigma _x}$ is the transverse rms beam size and $\left| \ell \right| + 1$ corresponds to the beam quality factor. This momentum spread should be large in comparison to the natural momentum spread of the electrons, which is at the cathode in the range of 0.2 and 0.4~keV/c for the typical kinetic emission energies of  0.2 - 0.5~eV~\cite{Ehberger2015}. Photo emitted electrons ($\sim$0.5~eV kinetic energy) are worse by a factor of 2-3 compared to field emitted electrons for typical photocathodes, but photo emission offers a much higher flexibility for shaping the transverse distribution and the size of the emitting area than field emission (the higher brightness of field emitted electron beams is primarily due to the small emission size). Lower transverse momenta can be reached by cryogenically cooled cathodes and by special photocathode materials as for example GaAs~\cite{Pastuszka2000}, which would also offer the possibility to produce spin polarized electrons.

If the transverse rms beam size of a ring shaped laser beam on the photocathode were in the micrometer range, $\left| \ell \right|$ would still need to be larger than $10^3$ to increase the divergence of the Laguerre-Gauss mode over the natural divergence of the electrons. According to Eq.~(\ref{Eqell}), the required magnetic field on the cathode would be in the range of 1~T in this case. The conditions of a pure mode, i.e., the momentum spread of the mode is larger than the natural momentum spread of the electrons, can be easier realized with lower magnetic fields and larger ring radii, if larger OAM values are acceptable. Deviations from the pure mode conditions due to a mismatch of the transverse shape or momentum lead, however, only to a population of neighboring modes, a beam with angular momentum is created in any case. 

The generation of pure modes with a low quantum number from an immersed cathode is challenging and would likely require a cryogenically cooled, structured photocathode, which delivers electrons with transverse momentum spread well below 0.2~keV/c and ideally small, high field solenoids ($B_{z,0} > $1~T), as can be generated by superconducting coils. Note that the high field values are only required in a very small volume with a transverse size of the order of the wave packet's width, so that the generation of significant fields is conceivable with reasonable effort. GaAs photocathodes would offer an additional opportunity to produce spin polarized electrons, however, with regard to surface preparation and vacuum conditions, the material is quite demanding. The generation of mixed modes with a large OAM bandwidth on the other hand is straightforward. Beams with a large OAM bandwidth can be useful for studying quantum entanglement, as discussed, for instance, in~\cite{Torres2003} and they appear also to be acceptable for some of the proposed applications of angular momentum beams. Some electron microscopes are already equipped with a photocathode laser, so that only the cathode geometry needs to be adapted and a solenoid needs to be installed. 

A big advantage of the immersed cathode technique -- besides its high efficiency -- is its high flexibility. For a fixed geometry of the ring-shaped emission only the magnetic field has to be controlled. 
The angular momentum can be freely adjusted according to Eq.~(\ref{Eqell}) and it can easily be reversed, which is mandatory for some of the proposed applications.
 
Above, the conditions at the cathode were discussed, while one would rather like to reach the ideal matching conditions in the fringe field region, where the angular momentum is imparted onto the beam. Optical imaging of the cathode plane into the plane of the fringe field might be necessary. Neither acceleration, nor a magnification of the cathode image will change the conditions concerning the uncorrelated transverse momenta, as discussed above. Thus, the use of the cathode in the solenoid field opens a simple route for the generation and manipulation of electron vortex beams with quantized angular momentum in electron microscopes.

{\it Other charged particles --}
The extension of the immersed cathode technique to other charged particles sources is -- disregarding technical limitations -- straight forward. For example it is conceivable to place the production target for positrons, anti-protons or other exotic particles into a solenoid. The poor beam quality of these particle sources will in general limit this approach to the generation of beams with broad OAM bandwidth. Better conditions are reached with Penning traps, where ultra-cold particles can be prepared, or with ion sources. In Electron Cyclotron Resonance (ECR) ion sources particles are for example ionized inside a solenoid field, the extracted ions carry thus naturally an OAM~\cite{Bertrand2006}. ECR sources are able to produce singly or multi charged ion beams and even radioactive~\cite{Bertrand2006} beams with high intensity. But the solenoid field strength is in this case fixed by the resonance condition and thus it is not a free parameter.

A flexible approach is to immerse a charge stripping medium, e.g., a stripping foil, into a solenoid~\cite{Groening2017, Groening2018}. It is common practice in ion accelerators to increase the charge state of the ions by passing them through a stripping medium. When the medium is immersed in a solenoid field the canonical angular momentum changes as (cf. Eq.~\ref{Eq1.6})
 \begin{equation}
\tilde{L} = \ell \hbar = e\frac{Z_{out}-Z_{in}}{2}B_{z,0}\langle r^2 \rangle
 \label{Eq1.7}
 \end{equation}
due to the stripping process. When the canonical angular momentum of the incoming state is zero, Eq.~\ref{Eq1.7} yields directly the kinetic angular momentum of the outgoing state.

In order to keep the degradation of the beam quality due to scattering in the stripping medium under control the beam has to be sufficiently focused, so that the incoming transverse momenta are increased~\cite{Groening2017, Hwang2014}. Thus a better beam quality of the incoming beam will require a smaller beam size in the stripping medium and correspondingly higher fields to create a certain angular momentum. A large variation of the charge state $\Delta Z= Z_{out}-Z_{in}$ on the other hand leads to relaxed requirments of the solenoid field strength. 

{\it Summary --} The quantum mechanical formulation of the Busch theorem is presented and its application for the generation of quantized charged vortex beams is discussed. The generation of beams with broad OAM bandwidth is straight forward and requires only a solenoid field of modest strength around the cathode -- for electrons -- or a charge stripping medium for the production of ions. In order to produce a pure mode with small OAM bandwidth in a microscope application a ring shaped cathode of small diameter and a high solenoid field is required in a small volume around the cathode.\\
The basic simplicity, the high efficiency and the great flexibility of the immersed source technique make it an attractive solution for the generation of quantized charged vortex beams of various charged particles from electrons to ions and exotic particles.

\begin{acknowledgments}
We are grateful to P.~Kazinski, V.~Serbo, and A.~Surzhykov for useful discussions.
\end{acknowledgments}

\end{document}